\documentclass[pdflatex,sn-mathphys-num]{sn-jnl}

\usepackage{adjustbox}
\usepackage{graphicx}%
\usepackage{multirow}%
\usepackage{amsmath,amssymb,amsfonts}%
\usepackage{amsthm}%
\usepackage{mathrsfs}%
\usepackage[title]{appendix}%
\usepackage{xcolor}%
\usepackage{textcomp}%
\usepackage{manyfoot}%
\usepackage{booktabs}%
\usepackage{algorithm}%
\usepackage{algorithmicx}%
\usepackage{algpseudocode}%
\usepackage{listings}%

\theoremstyle{thmstyleone}%
\theoremstyle{thmstyletwo}%

\theoremstyle{thmstylethree}%

\begin{document}

\title[Assessing Extrapolation of Peaks Over Thresholds with Martingale Testing]{Assessing Extrapolation of Peaks Over Thresholds with Martingale Testing}
\subtitle{Application to EVA2025 Data Challenge}


\author[1]{\fnm{Joseph} \sur{de Vilmarest}}\email{joseph.de-vilmarest@vikingconseil.fr}

\author[2]{\fnm{Olivier} \sur{Wintenberger}}\email{olivier.wintenberger@sorbonne-universite.fr}

\affil[1]{\orgname{Viking Conseil}, \orgaddress{\street{26 rue de la Pie}, \postcode{28000} \city{Chartres}, \country{France}}}

\affil[2]{\orgdiv{LPSM}, \orgname{Sorbonne Université}, \orgaddress{\street{4 place Jussieu}, \postcode{75005} \city{Paris}, \country{France}}}


\abstract{We present the winning strategy for the EVA2025 Data Challenge, which aimed to estimate the probability of extreme precipitation events. These events occurred at most once in the dataset making the challenge fundamentally one of extrapolating extreme values.
Given the scarcity of extreme events, we argue that a simple, robust modeling approach is essential. We adopt univariate models instead of multivariate ones and model Peaks Over Thresholds using Extreme Value Theory. Specifically, we fit an exponential distribution to model exceedances of the target variable above a high quantile (after seasonal adjustment).
The novelty of our approach lies in using martingale testing to evaluate the extrapolation power of the procedure and to agnostically select the level of the high quantile. While this method has several limitations, we believe that framing extrapolation as a game opens the door to other agnostic approaches in Extreme Value Analysis.}




\maketitle

\section{Introduction}

The EVA2025 Data Challenge was held in advance of the Extreme Value Analysis (EVA) Conference in June 2025. The challenge focused on estimating the frequencies of occurrence of three extreme precipitation events across 50 runs of a climate model, based on only 4 runs accessible to the competitors. Teams were evaluated based on the accuracy of their estimated frequencies and associated confidence intervals. The targeted events are present only once (event 3) or even absent (events 1 and 2) in the 4 available runs. Therefore, the challenge consists in extrapolating extreme values.

Facing the challenges of extrapolation, we classify potential methods into three distinct approaches.

The class of AI for Extremes has been developed recently, fitting deep generative neural networks to the entire dataset and trusting the generation of extremes from the model. This strategy has been applied successfully to meteorological forecasting; it relies on no assumptions and could be transposed to our framework, thanks to the universality of neural networks. However, the extrapolation power of such approaches remains questionable \cite{zhang2025numerical}.

The class of Machine Learning for EVA adapts learning methods to exceedances of a high threshold. The topic is more mature thanks to recent developments in the EVA community. Fitting complex multivariate models requires a sufficient amount of data and therefore choices of moderate thresholds. Validating the extrapolation power of such an approach remains challenging and usually coincides with validating the choice of the model used for the angular distribution of the exceedances \cite{kiriliouk2019peaks}. Given the high level of the targeted events in the Data Challenge, we argue that a good model above a moderate threshold might not provide accurate frequencies.

The last class corresponds to the application of the foundations of Extreme Value Theory. This approach is simple and its extrapolation power is guaranteed under assumptions widely considered reasonable. The main restriction is that the theory is mostly univariate, i.i.d. \cite{de2006extreme}, and relies on the existence of an unknown threshold, potentially large, that must be estimated. Given the scarcity of extreme events targeted in the challenge, we argue that such a simple modeling approach of exceedances above high thresholds is essential.

As discussed in the introductory article of the Data Challenge, we did not find evidence against stationarity. However, a strong seasonality in the extremes is observed, leading us to model the day-of-year variable and its influence on the exceedances. We apply our approach considering univariate targets as i.i.d. after seasonal adjustment. The three targets are events relative to daily precipitations on a 5$\times$5 grid of 25 locations. The first two targets are order statistics of the 25 precipitations of a given day; therefore, we reduce the 25-dimensional problem to a univariate one by considering the relevant order statistics. The third target combines order statistics of the 25 precipitations of two consecutive days. We formulate this task as a 2-dimensional problem by considering the daily couples of relevant order statistics; then, we apply our univariate approach to an auxiliary problem using an adequate norm. We could have transformed Target 3 into a univariate one directly. However, we would have missed a strong pattern observed in the extreme order statistics of two consecutive days.

Our univariate procedure follows the classical Peaks Over Threshold (POT) methodology \cite{coles}. We fit a nonlinear seasonal model on the scale parameter of the exceedances, fixing the shape parameter to $0$. This yields seasonally adjusted targets above a high quantile that we model by a 1-mean exponential distribution.

The only delicate point of the POT approach is the threshold selection. We introduce a novel approach to assess the quality of extrapolation of a model by martingale testing by betting \cite{shafer2021testing}. We define the sequential game of extrapolation as predicting the next top order statistics given previous ones. To assert extrapolation, we compare the predictions of the model in this game with the observations. The difference between observed top order statistics and sampled ones from the model constitutes a martingale if the model is coherent with the observations in the extrapolation game. To provide an agnostic score of extrapolation, we rely on the terminal wealth of an adversary that bets on the sign of the next error of prediction.
We choose thresholds minimizing this score. Therefore, the chosen threshold can be seen as the one whose signs of errors on top-order statistics prediction are the least predictable.
Relying on this threshold selection led us to win the data challenge, and we believe it might be of interest for various domains. Our implementation is available on gitlab\footnote{Implementation: \url{https://gitlab.com/JdeVilmarest/eva2025-data-challenge-viking-sorbonne}}.

We present the data challenge as well as our model in Section~\ref{sec:pot}. The threshold selection approach relying on martingale testing is introduced in Section~\ref{sec:testing}, then applied to our use case in Section~\ref{sec:practice}.

\section{POT Models for Extrapolation}\label{sec:pot}
The challenge focuses on predicting extreme precipitation events in a dataset of climate models. Specifically, it involves estimating the frequency of rare events over 50 climate runs using only 4 available runs; each run lasts 165 years, that is $165\times 365$ data points. The objective is to estimate the frequency of these events and their confidence intervals, with evaluation based on comparing the true frequencies from the 50 runs with the forecasts. Notably, the empirical frequencies of the three targets on the 4 available runs are $0$, $0$, and $\frac{1}{4 \times 165 \times 365} \approx 5 \times 10^{-6}$, making this fundamentally a problem of extrapolation.

\subsection{From Multivariate to Univariate}\label{sec:univariate}
Given the scarcity of the targets, we advocate for simple models. Although the problem initially involves multivariate modeling of a 5$\times$5 grid of precipitation data, we reduce each target to a univariate problem. Below, we present our univariate formulation for the three initial targets, concatenating the 4 runs and letting $t$ range from $1$ to $4 \times 165 \times 365$.

\begin{figure}[!]
\centering
\includegraphics[width=0.325\textwidth]{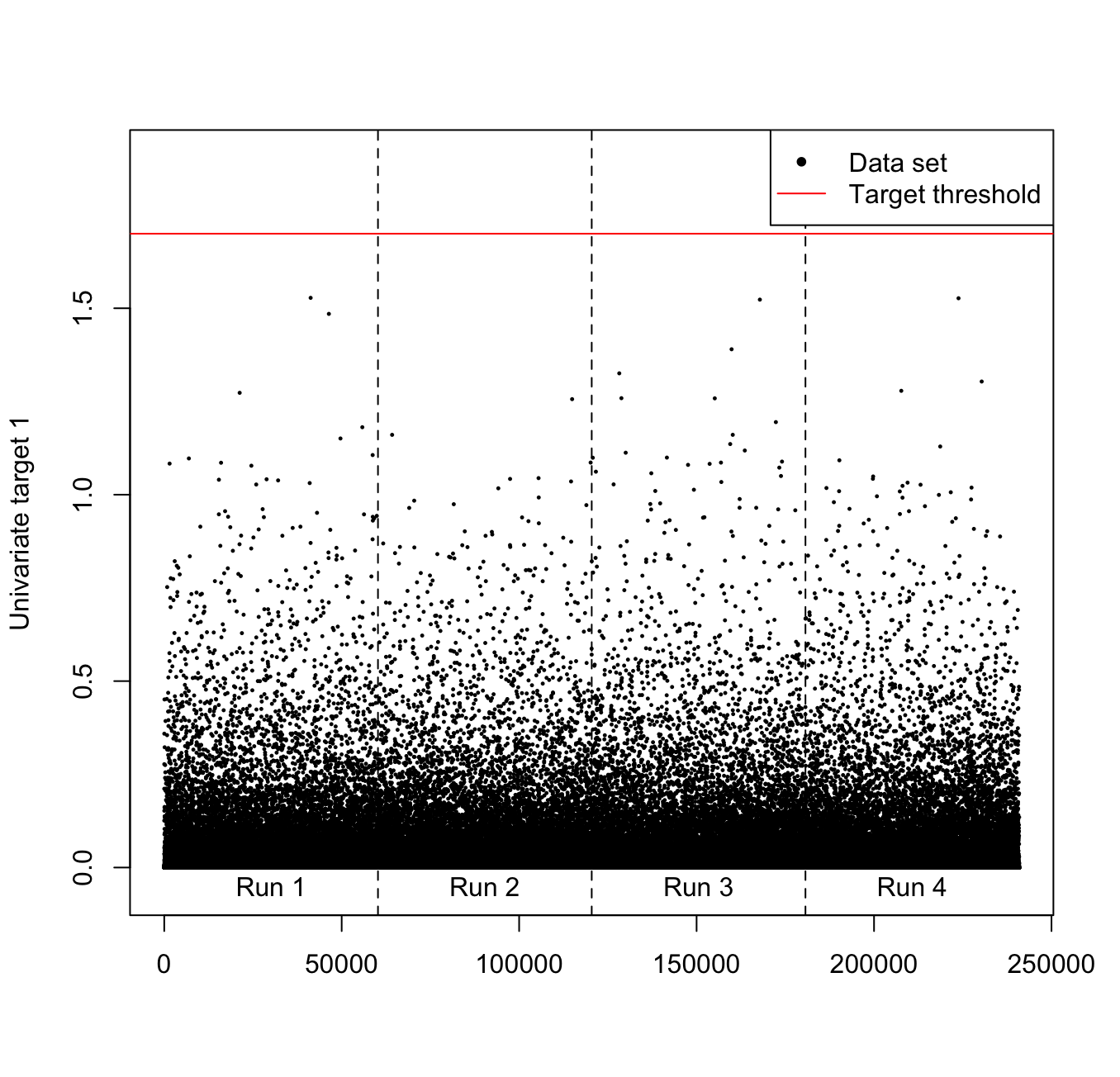}
\includegraphics[width=0.325\textwidth]{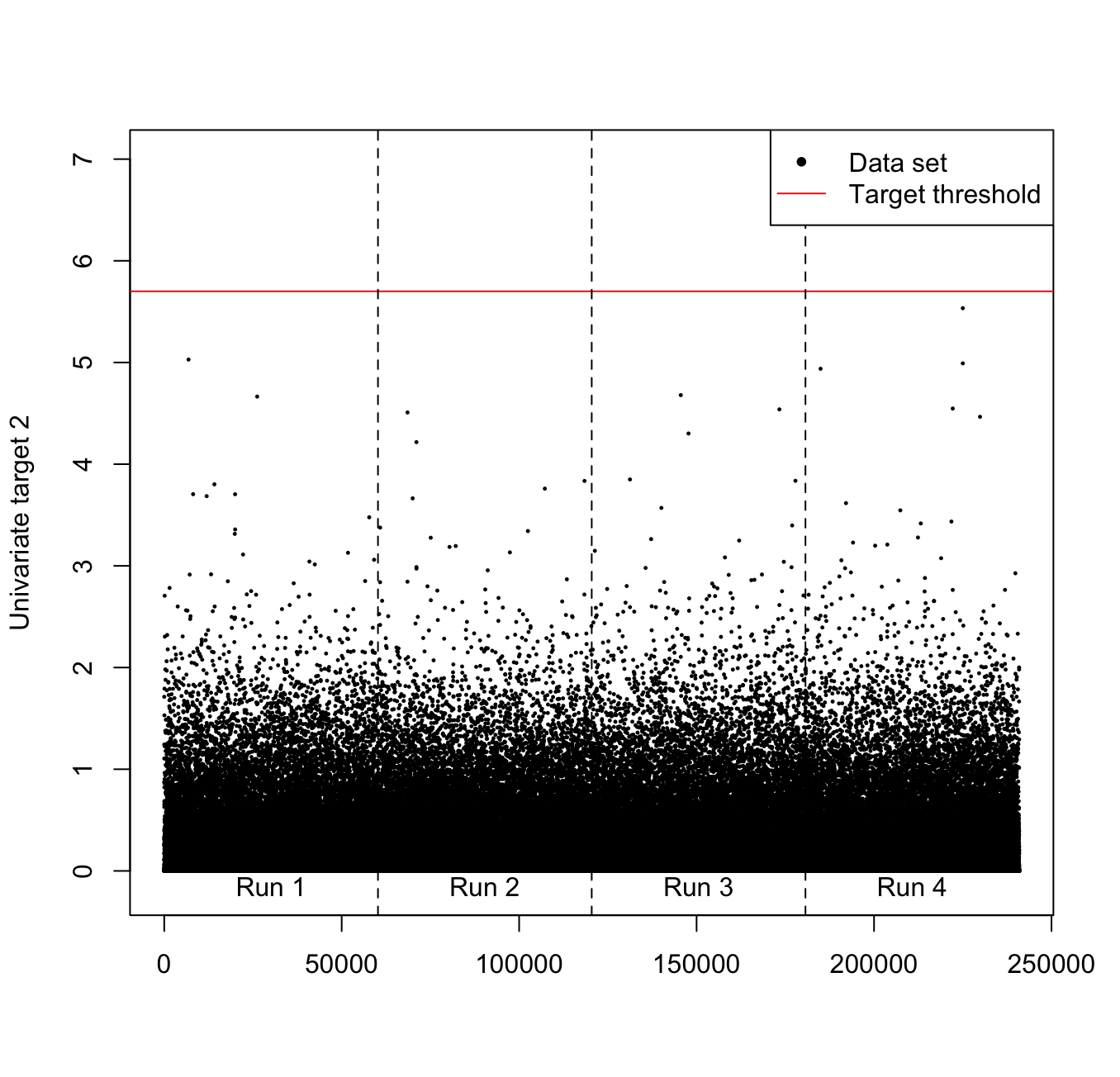}
\includegraphics[width=0.325\textwidth]{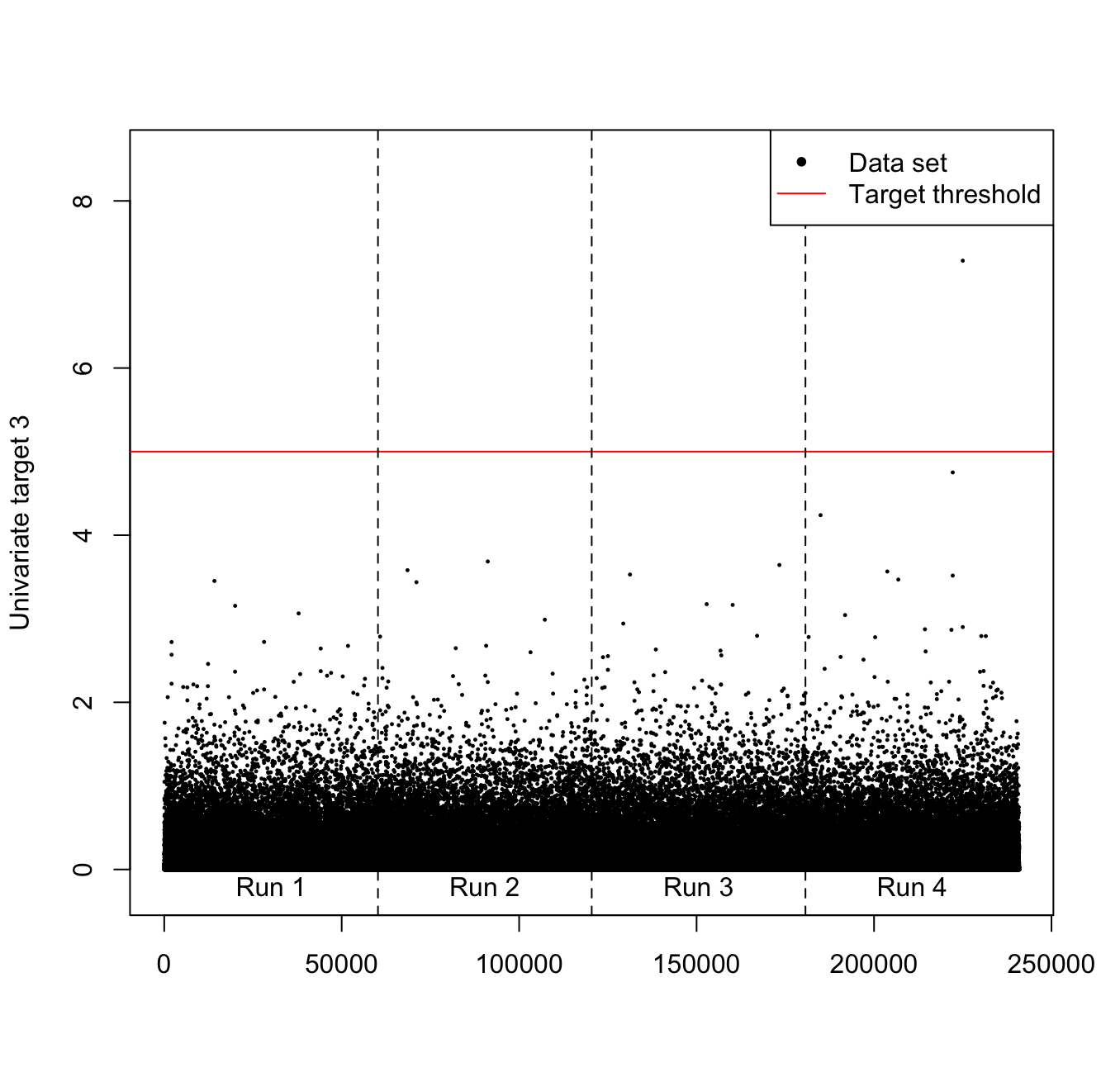}
\caption{Univariate targets concatenated on the 4 available runs of 165 years.}\label{fig:targets_univariate}
\end{figure}

\begin{figure}[!]
\centering
\includegraphics[width=0.4\textwidth]{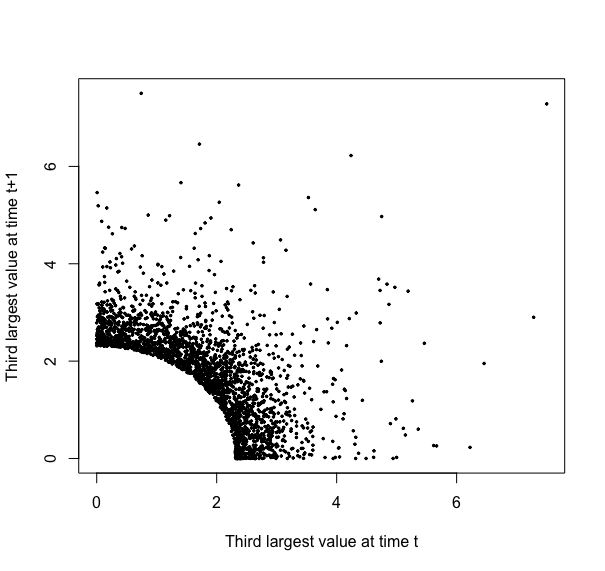}
\includegraphics[width=0.4\textwidth]{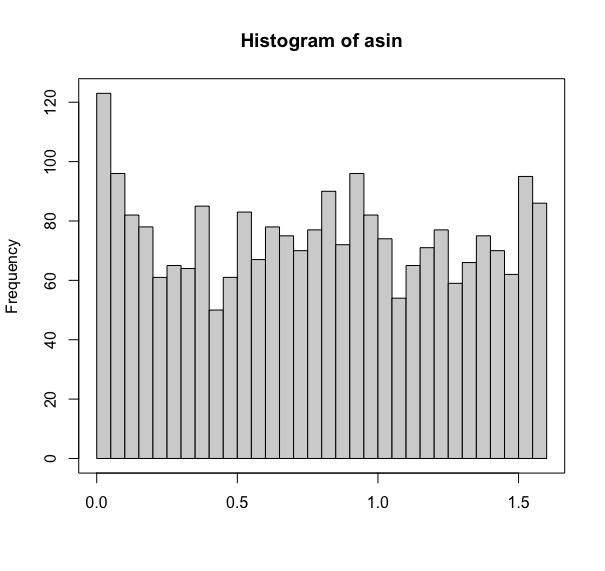}
\caption{Analysis of $(y_t^{(3.1)}, y_t^{(3.2)}) \mid \overline{y}_t^{(3)} > u$ where $u$ is the quantile 0.99 of $ \overline{y}_t^{(3)}$. On the left: $y_t^{(3.2)}$ as a function of $y_t^{(3.1)}$. On the right: histogram of $\arcsin(y_t^{(3.1)}/\overline{y}_t^{(3)})$.}\label{fig:target3_aux}
\end{figure}

\begin{itemize}
\item
\textbf{Target 1} is the expected number of times per model run that all 25 locations exceed 1.7 Leadbetters. We define $y_t^{(1)}$ as the minimum value across the 25 locations at time $t$, with threshold $T_1 = 1.7$.

\item
\textbf{Target 2} is the expected number of times per model run that at least 6 of the 25 sites exceed 5.7 Leadbetters. We define $y_t^{(2)}$ as the 6th largest value across the 25 locations at time $t$, with threshold $T_2 = 5.7$.

\item
\textbf{Target 3} is the expected number of times per model run that at least 3 of the 25 sites exceed 5 Leadbetters for a run of at least two consecutive days. We define $y_t^{(3.1)}$ (resp. $y_t^{(3.2)}$) as the 3rd largest value across the 25 locations at time $t$ (resp. $t+1$), then $y_t^{(3)}$ as $\min(y_t^{(3.1)}, y_t^{(3.2)})$, with threshold $T_3 = 5$.
\end{itemize}
Our univariate formulation for Target $i$ is the expected number of times per model run that $y_t^{(i)} \ge T_i$. Note that our definition of Target 3 slightly differs from the initial one of the organizers; indeed, if $y_t^{(3)}>5$ and $y_{t+1}^{(3)}>5$, then we would count two occurences, while the organizers count only one. We do that approximation because of our i.i.d. assumption presented in Section~\ref{sec:pot_modeling}.
The three targets, along with their respective thresholds, are displayed in Figure~\ref{fig:targets_univariate}.
In addition, defining
$
\overline{y}_t^{(3)} = \sqrt{(y_t^{(3.1)})^2 + (y_t^{(3.2)})^2},
$
it can be observed in Figure~\ref{fig:target3_aux} that for a sufficiently large $u$, the conditional distribution of
\[
(\overline{y}_t^{(3)})^{-1}(y_t^{(3.1)}, y_t^{(3.2)}) \mid \overline{y}_t^{(3)} > u
\]
is close to being uniformly distributed on the positive quadrant of the unit sphere; we therefore model it as $(\sin\theta, \cos\theta)$, where $\theta \sim \mathcal{U}([0, \pi/2])$. Therefore, our model for  Target 3 is structured in two steps:
\begin{enumerate}
    \item We model the extremes of $\overline{y}_t^{(3)}$.
    \item We model $
    y_t^{(3)} = \overline{y}_t^{(3)} \min(\sin\theta_t, \cos\theta_t)$,
    where $(\theta_t)$ is a sequence of i.i.d. realizations from $\mathcal{U}([0, \pi/2])$.
\end{enumerate}
In summary we simplify the framework by defining a univariate target for each task and estimate each frequency independently.

\subsection{Extrapolated Values in POT Modeling}\label{sec:pot_modeling}
Following classical extreme value analysis, we consider exceedances above a high quantile  $q$. Despite the dynamic nature of climate runs, no clear trend is observed in the data. We thus assume that $(y_t^{(i)})_t$ is a sequence of independent realizations of a unique random variable $Y^{(i)}$, embedding our work in classical Extreme Value Theory; it allows to simply concatenate the runs, attributing the same weight to each day.

It is well-known \cite{bdh,pickands1975} that exceedances of a random variable $Y^{(i)}$ over a large threshold $q_i$ are well approximated by a Generalized Pareto Distribution (GPD) $GP_{\sigma, \gamma}(x)=  1 - \left(1 + \gamma \frac{x}{\sigma}\right)_+^{-1/\gamma}$ for some $\sigma>0$, $\gamma \in \mathbb{R}$, i.e.,
\begin{align*}
    Y^{(i)}-q_i \mid Y^{(i)}\ge q_i \sim GP_{\sigma, \gamma}\,.
\end{align*}

In the POT approach, we select a sufficiently large quantile $q_i$ to fit a GPD model to the exceedances $Y^{(i)} - q_i$ given $Y^{(i)} > q_i$, relying on the fitted model for extrapolation. The extrapolation is particularly sensitive to the estimation of the shape parameter $\gamma$; we simplify and stabilize the extrapolation procedure by fixing $\gamma=0$, that is we recover the exponential distribution. Figure \ref{fig:qqplot_E} shows that the fit of this simple model is reasonable after seasonal adjustment.

\subsection{Seasonal Adjustment}
\begin{figure}
\centering
\includegraphics[width=0.325\textwidth]{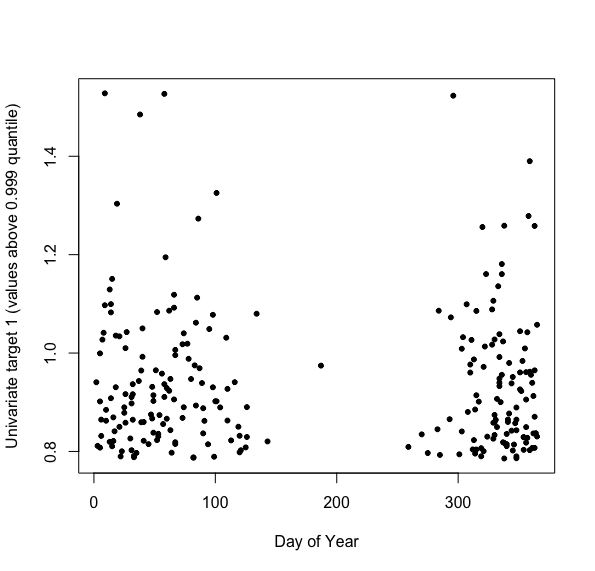}
\includegraphics[width=0.325\textwidth]{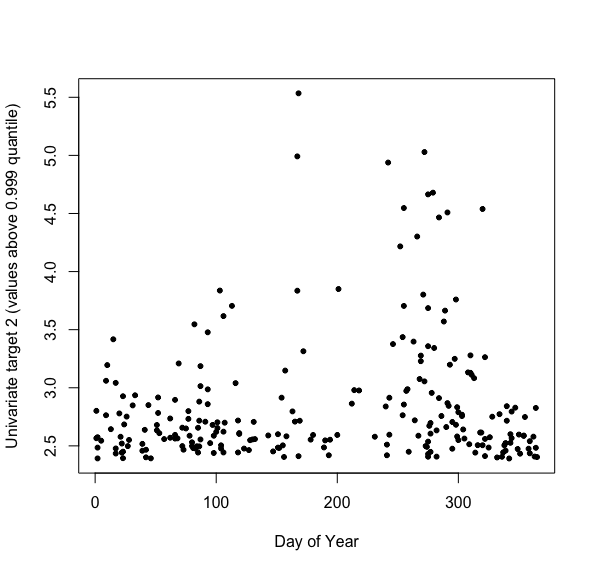}
\includegraphics[width=0.325\textwidth]{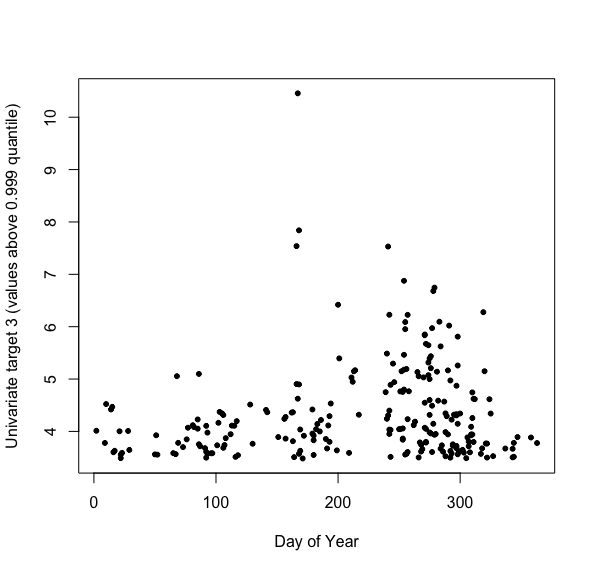}
\caption{Analysis of each univariate target as a function of the time of year. Here $p_i=0.999$ for $i=1,2,3$.}\label{fig:posan_target}
\end{figure}

As mentioned previously, we did not find evidence against stationarity. However, there is a strong seasonal effect, particularly in the extremes. We display in Figure~\ref{fig:posan_target} the dependence of the targets to the day of year, ranging from 1 on January 1\textsuperscript{st} to 365 on December 31\textsuperscript{st}, and denoted by $d_t$ for any $t\in\{1\hdots 4\times 165\times 365\}$.
For any $i\in\{1,2,3\}$, we apply the POT approach to the target $Y^{(i)}_t$ given that $d_t=d$. 
We take into account the influence of $d$ in the scaling factor $\sigma_{i,d}=f_i(d)>0$. For each value of $d$ and $i$ we approximate:
\begin{align*}
    & \mathbb P(Y^{(i)}_t-q_i  \leq x \mid Y^{(i)}_t> q_i, d_t=d) \approx GP_{f_i(d),0}(x) \,.
\end{align*}
Remark that $f_i(d)E\sim
GP_{f_i(d),0}$ for $E$ standard exponentially distributed.
Also, we observe on Figure~\ref{fig:posan_target} that the day of year does not only affect the value of the exceedances but also their frequency of appearance. Our model accounts for that using the empirical distribution of the day of year across exceedances.
Therefore, we approximate the distribution of the exceedances with two auxiliary random variables: a random day $D$ distributed uniformly from all days
on which the observed value exceeds the threshold, a seasonally adjusted exceedance $E$ following the exponential distribution:
$$
\begin{cases}
&D \sim \mathcal{U}(\{ d_t \mid y_t^{(i)} > \hat F_i^{-1}(p_i)\})\,,\\
&Y^{(i)} - \hat F_i^{-1}(p_i) \approx  f_i(D) \times E\,,
\end{cases}
$$
where $q_i=\hat F_i^{-1}(p_i)$ is the empirical $p_i$-quantile of $y_t^{(i)}$, $f_i$ is a nonlinear effect to be estimated, and $E$ is a random variable standard exponentially distributed.

\begin{figure}
\centering
\includegraphics[width=0.325\textwidth]{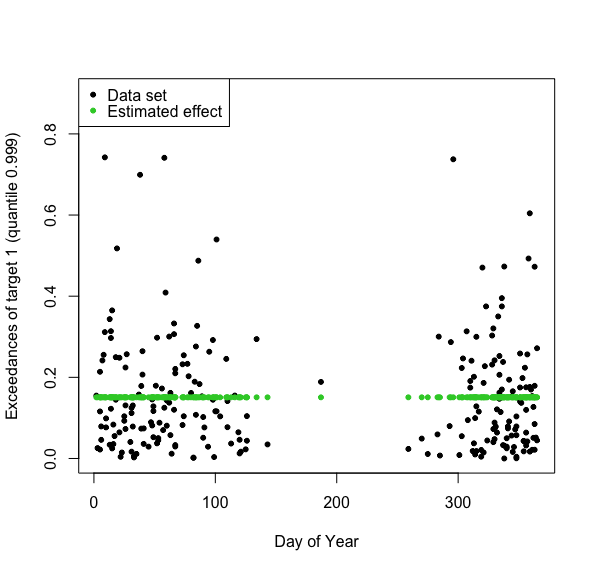}
\includegraphics[width=0.325\textwidth]{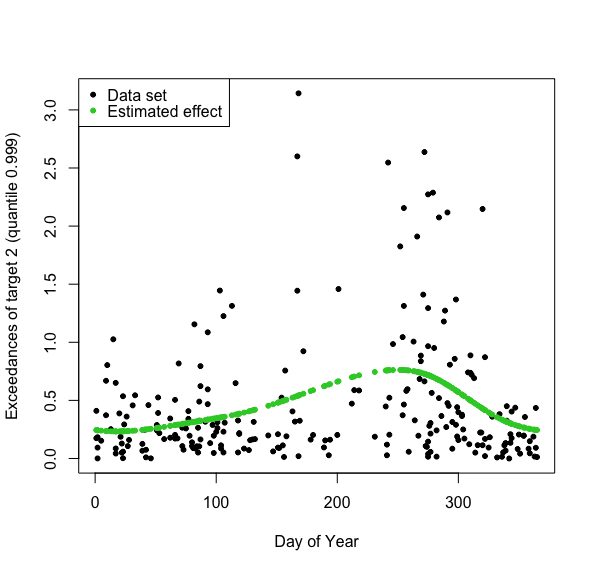}
\includegraphics[width=0.325\textwidth]{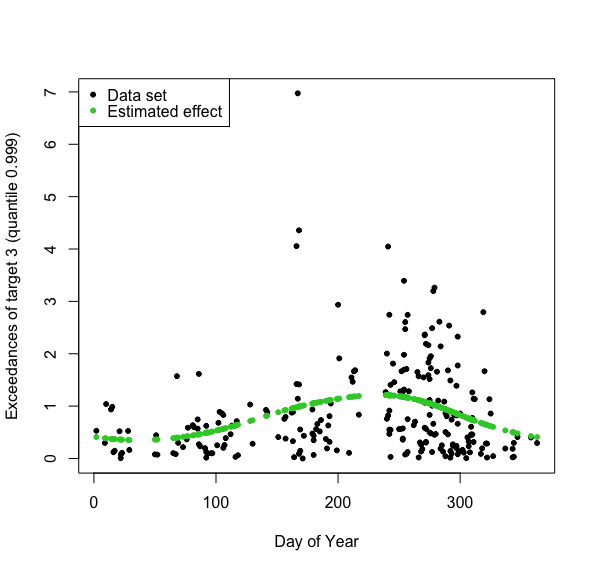}
\includegraphics[width=0.325\textwidth]{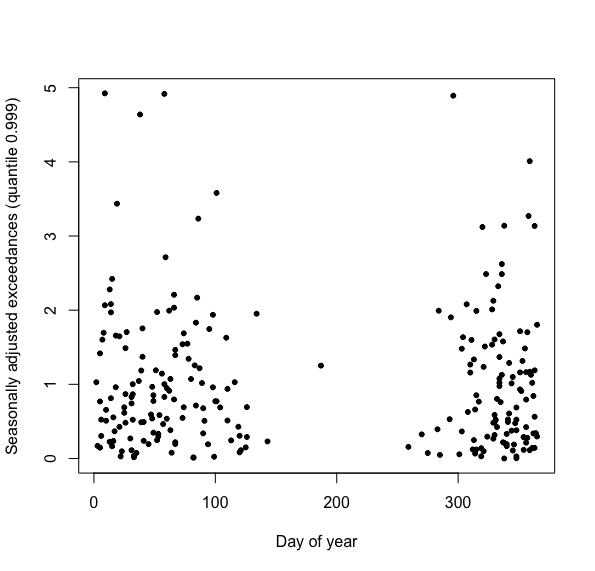}
\includegraphics[width=0.325\textwidth]{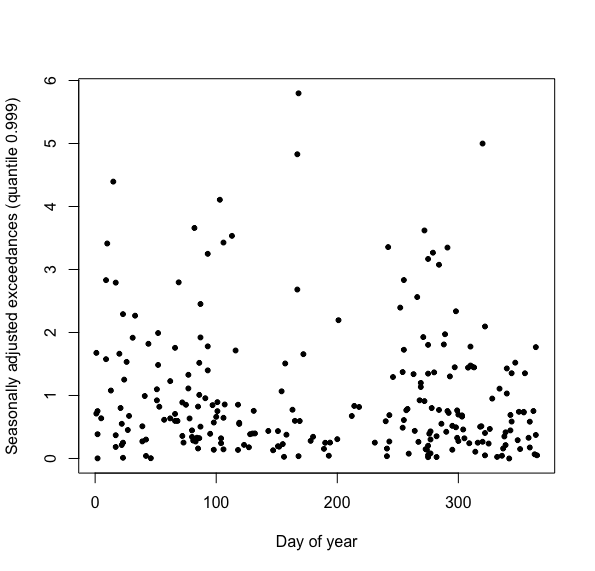}
\includegraphics[width=0.325\textwidth]{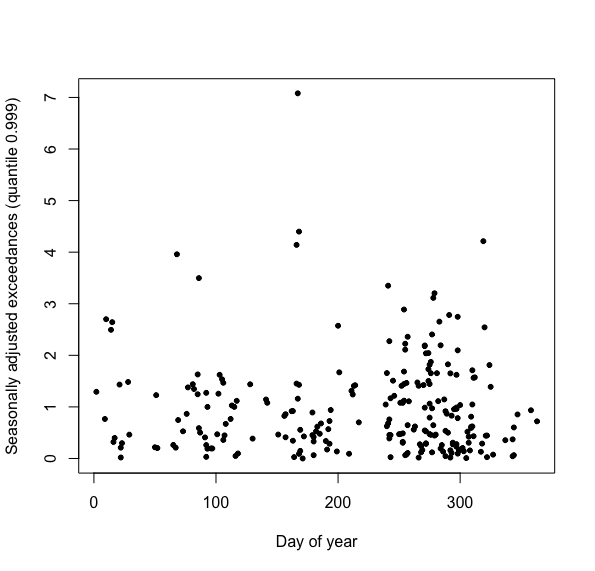}
\caption{Top: exceedances depending on the day of year, as well as seasonal effects. Bottom: seasonally adjusted exceedances $E_t^{(i)}$. Here $p_i=0.999$ for $i=1,2,3$.}\label{fig:seasonal_effects}
\end{figure}

We estimate $f_i$ as a linear combination of pre-defined cyclic cubic splines, using the R library \texttt{mgcv} \citep{wood2001mgcv} with the quadratic loss function. The resulting $f_i$, as well as the seasonally adjusted exceedances $e_t^{(i)} =(y_t^{(i)}-\hat F_i^{-1}(p_i)) / f_i(d_t)$ are displayed in Figure~\ref{fig:seasonal_effects}.

\begin{figure}
\centering
\includegraphics[width=0.325\textwidth]{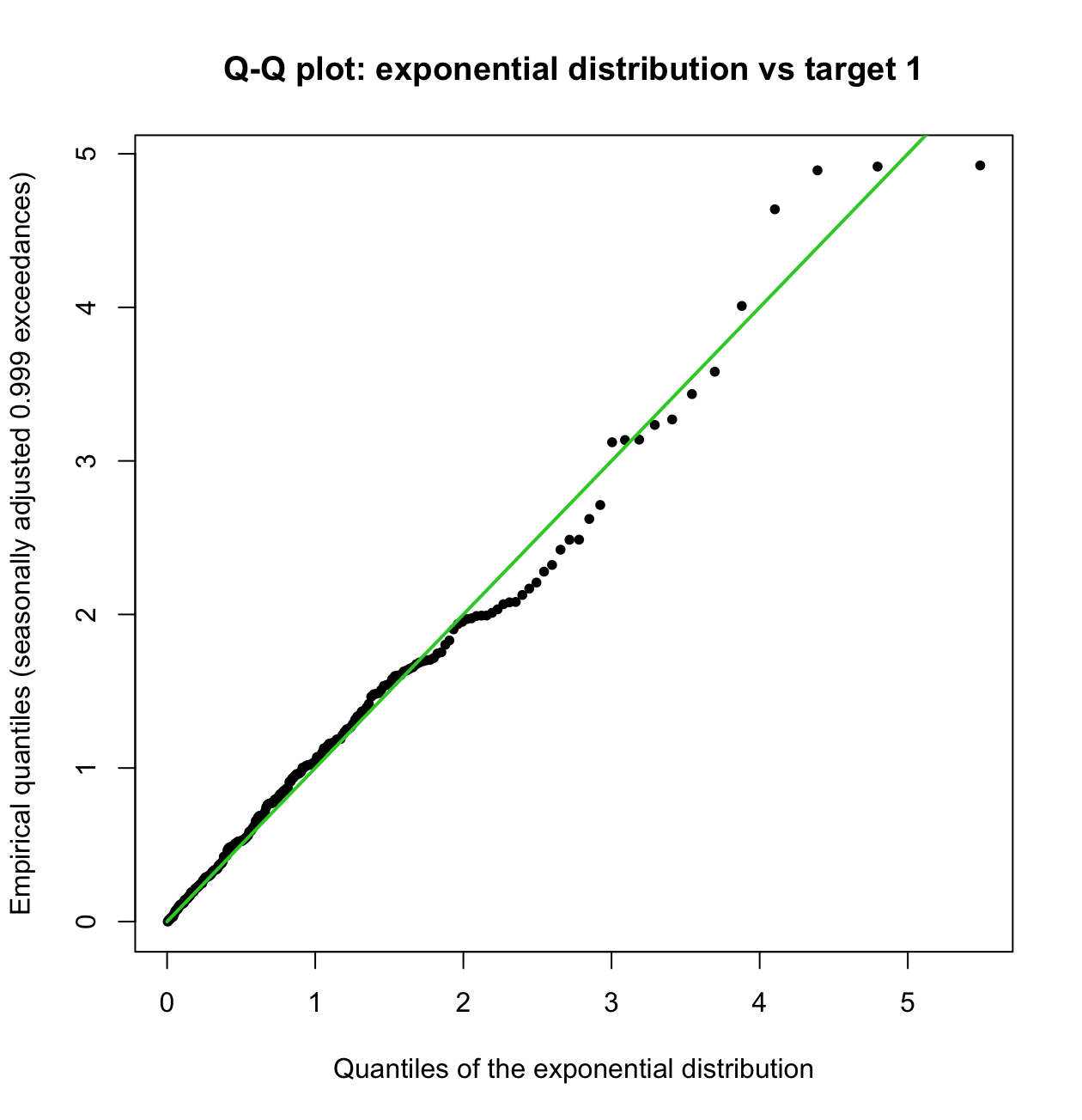}
\includegraphics[width=0.325\textwidth]{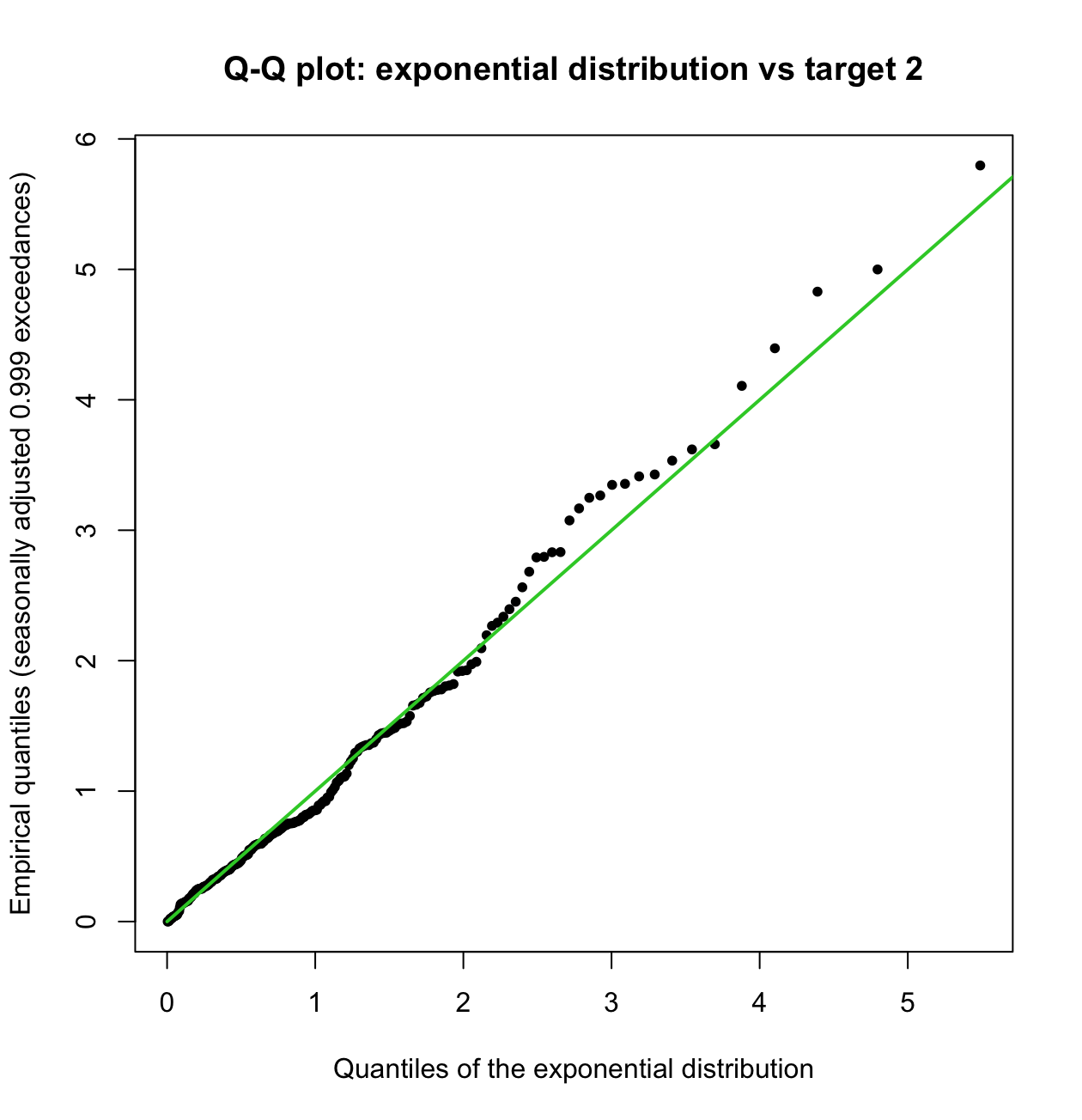}
\includegraphics[width=0.325\textwidth]{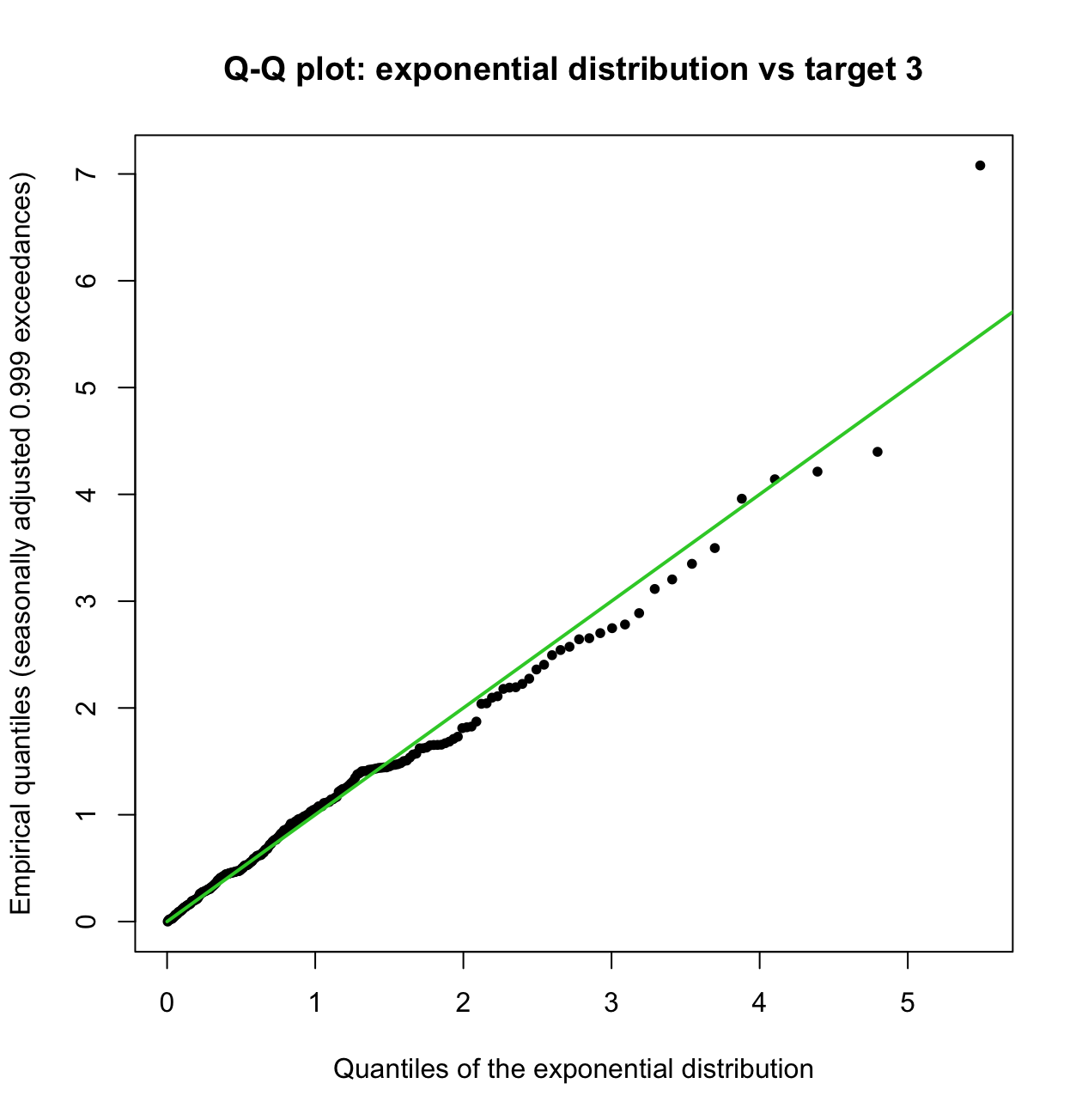}
\caption{Q-Q plot of the seasonally adjusted 0.999 exceedances, with the exponential distribution.}\label{fig:qqplot_E}
\end{figure}

Our obtained effects are dependent on the task, and we observe that $f_1$ is constant, that is we learn no seasonality on the exceedances above the 0.999 quantile.
We confirm graphically that the seasonally adjusted exceedances are independent of the day of year. Also, we observe in Figure~\ref{fig:qqplot_E} that their distributions resemble the exponential distribution.

\subsection{Extrapolation Pipeline}
We present our method with a key parameter, the quantile levels $p_i$, that needs to be chosen (that is the object of Section~\ref{sec:testing}).
\begin{enumerate}
\item
We use the 4 available runs to fit a POT model on the exceedances of our univariate target $y_t^{(i)}$ above its empirical $p_i$ quantile $\hat F_i^{-1}(p_i)$. We fit a cyclic nonlinear effect of the day of year on exceedances, yielding seasonally adjusted exceedances that we assume independent and exponentially distributed. Our simple POT model is:
\begin{equation}\label{eq:generative}
\begin{cases}
&D \sim \mathcal{U}(\{ d_t \mid y_t^{(i)} > \hat F_i^{-1}(p_i)\})\,,\\
&\Tilde Y^{(i)} = \hat F_i^{-1}(p_i) + f_i(D) \times E\,,
\end{cases}
\end{equation}
where $D$ and $E$ are independent random variables. It is critical to note that $D$ follows the empirical distribution of the day of year on the exceedances and not the uniform distribution. $E$ follows the standard exponential distribution.\\
For the third task, as described in Section~\ref{sec:univariate}, we model the exceedances of the auxiliary univariate target $\overline{y}_t^{(3)}$ above its empirical $p_3$-quantile $\hat F_3^{-1}(p_3)$ as in \eqref{eq:generative}. Then, the distribution of the initial target given the auxiliary target exceeds $\hat F_3^{-1}(p_3)$ is estimated by the one of the random variable\footnote{
The exceedances of the initial target do not exactly match the ones of the auxiliary target; nonetheless, $Y^{(3)}> 5$ implies $\overline{Y}^{(3)}> \sqrt{50}$, therefore as long as $u<\sqrt{50}$ we have $\mathbb{P}(Y^{(3)}> 5)=\mathbb{P}(Y^{(3)}> 5\mid \overline{Y}^{(3)} > u)\mathbb{P}(\overline{Y}^{(3)} > u)$. We ensure that the selected quantile is such that $\hat F_3^{-1}(p_3)<\sqrt{50}$.}
\begin{equation}\label{eq:generative3}
\Tilde Y^{(3)} = (\hat F_3^{-1}(p_3) + f_i(D) E) \min(\sin\Theta, \cos\Theta) \,,
\end{equation}
where $D$ and $E$ are independent of $\Theta\sim\mathcal{U}([0,\pi/2])$.

\item 
We use \eqref{eq:generative} and ~\eqref{eq:generative3} to assess the number of occurrences of the events of interest in the 46 runs of the climate model that were not provided.
Precisely, we apply two steps to define our estimated frequency and its associated confidence interval:
\begin{itemize}
\item
We use the preceding POT model as a way to generate under the univariate target's estimated distributions. We generate $\lceil (1-p_i) \times 46 \times 165 \times 365 \rceil$ i.i.d. realizations of the random variable $\Tilde Y^{(i)}$ and count the number of times the  event of interest occurs. Adding $0$ (resp. $1$) for tasks 1 and 2 (resp. 3) due to the 4 given runs yields a number of occurrences of the initial event, and dividing by $50$ yields a frequency of exceedances per run. Then, applying a Monte Carlo method, we repeat this 
experiment $N=10^4$ times independently. The point estimate we finally submitted is the median of the $N$ frequencies of exceedances per run. By construction it is a multiple of 0.02 as the target. 
\item
For confidence intervals, we first define a mean frequency of exceedances per run $m_i, i\in\{1,2,3\}$.
We rely on the Poisson approximation of rare events, see for instance Chapter 1 of \cite{mikosch2024extreme}. Indeed, if we had $n$ i.i.d. realizations $(Y^{(i)}_k)_{1\le k\le n}$ of a target random variable $Y^{(i)}$, then the Poisson approximation
\[
\sum_{k=1}^n 1(Y^{(i)}_k > u_i ) \sim \text{Poiss}(n\mathbb P(Y^{(i)} > u_i))\,,
\]
holds under the assumption that $n\mathbb P(Y^{(i)} > u_i)$ converges to a positive constant as $u_i\to \infty$ together with $n\to \infty$.
In our case, we make the assumption that $n=50 \times 165 \times 365$ and the threshold $u_i=T_i$ are sufficiently large given that $m_i < 1$ for each $i$. Next we estimate $n\mathbb P(Y^{(i)} >T_i)$ by the mean number of occurrences $50\,m_i$.
The histograms of the Poisson distributions are displayed in Figure~\ref{fig:poisson}. 
\begin{figure}
\centering
\includegraphics[width=0.325\textwidth]{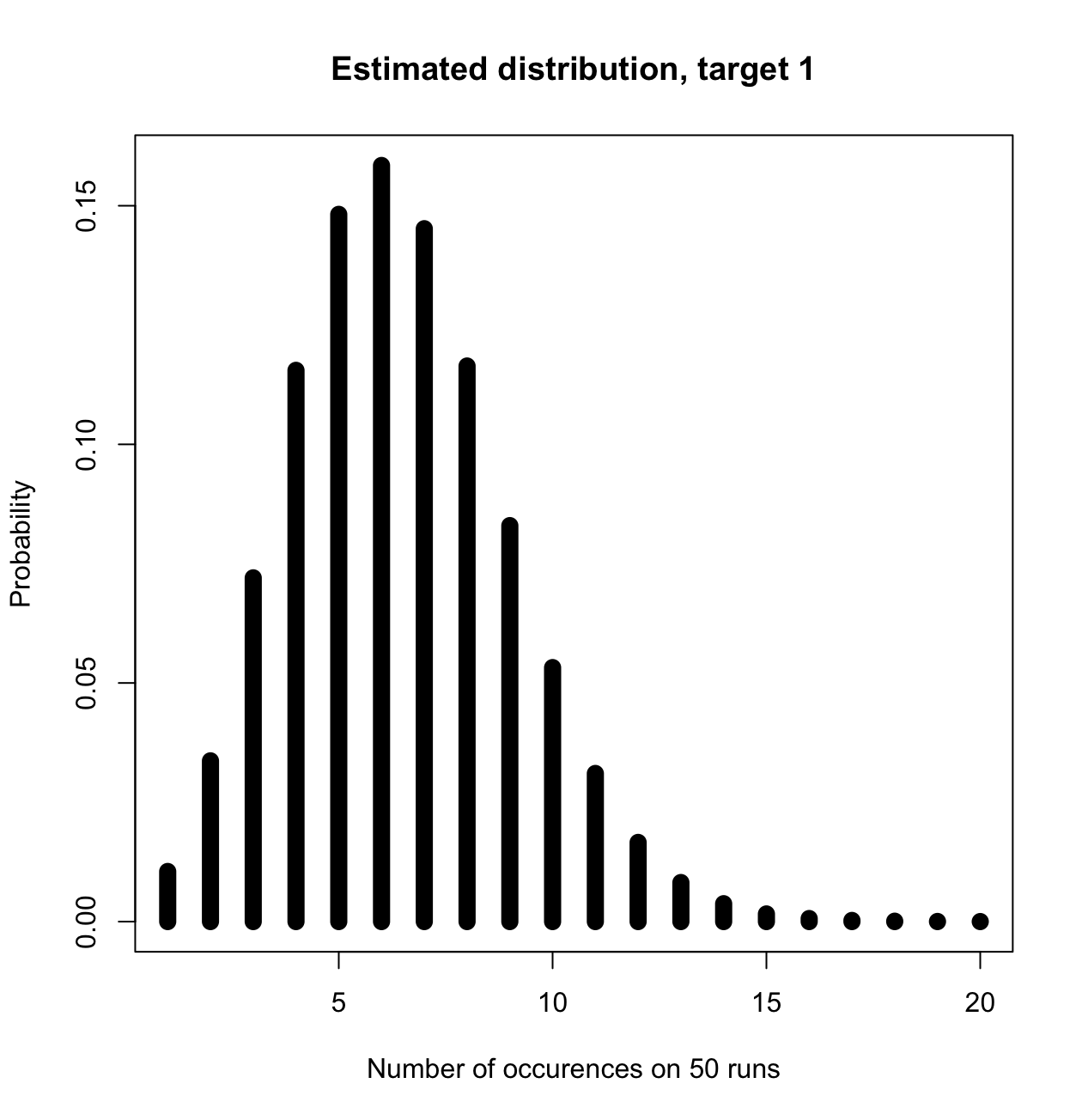}
\includegraphics[width=0.325\textwidth]{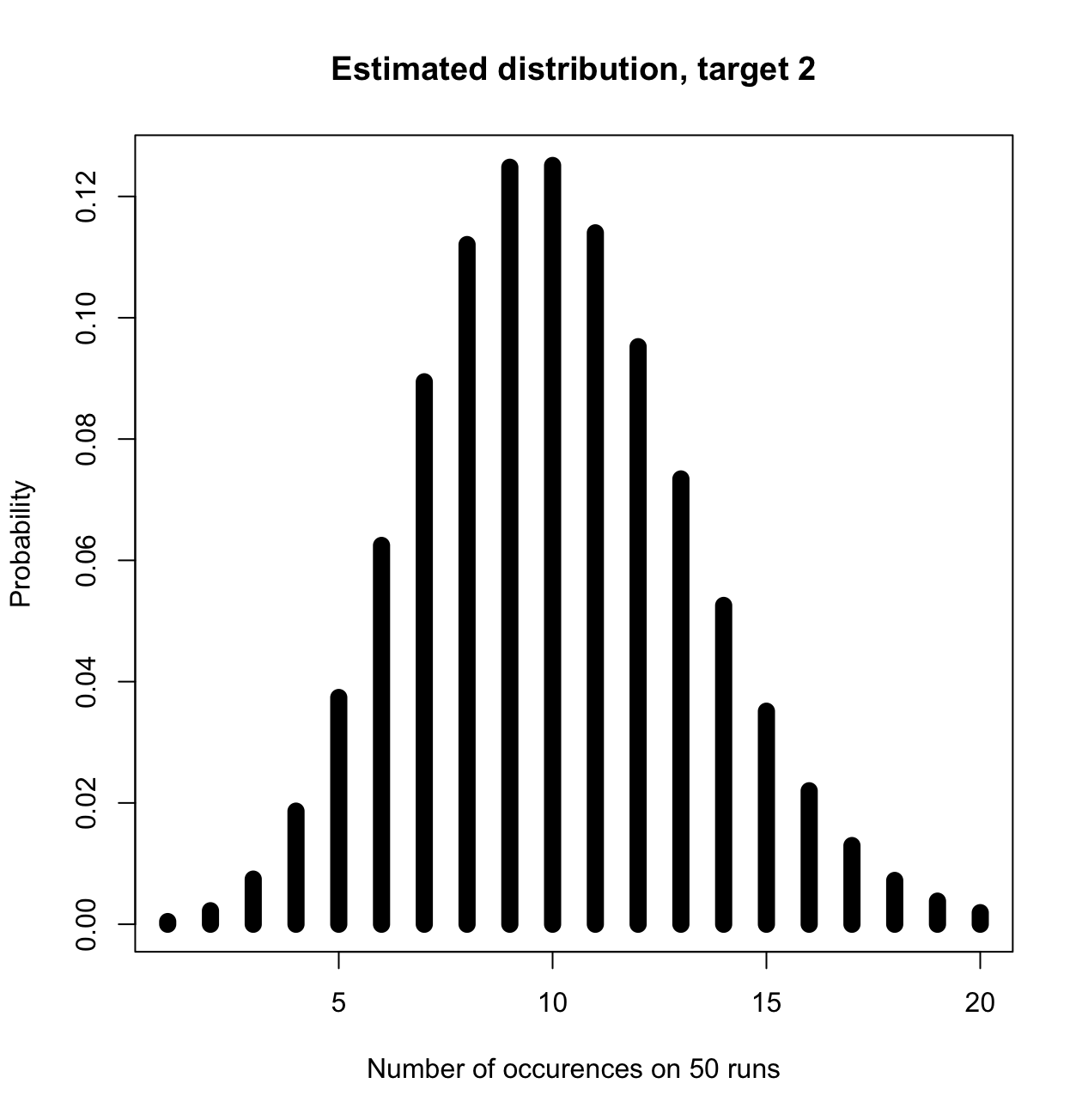}
\includegraphics[width=0.325\textwidth]{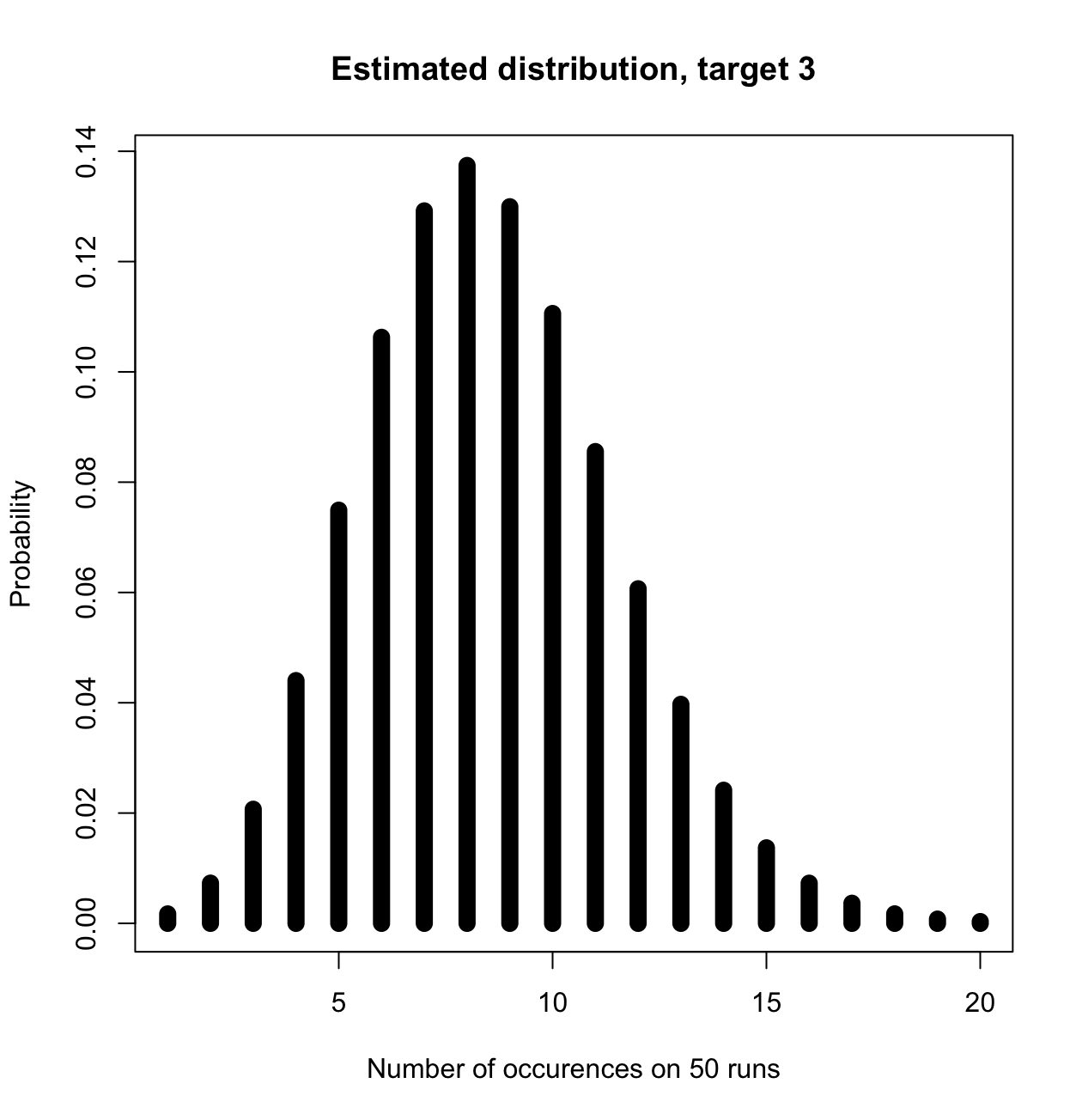}
\caption{Final Poisson distributions $Poiss(50\,m_i)$ used for the confidence intervals.}\label{fig:poisson}
\end{figure}
We chose the minimal-length interval achieving the desired confidence, and used confidence intervals of around 92\% instead of 95\% to improve the sharpness of our intervals.
It yields a confidence interval for the number of occurrences in the 50 runs, that we divide by 50 to obtain a confidence interval on the frequency per run.
As the scoring rule was a function only of the true frequency being in the interval and the length of the interval, it was useful to not consider bounds that are no multiple of 0.02.
\end{itemize}
\end{enumerate}

\section{Martingale Testing of Extrapolation Power}
\label{sec:testing}
Recall that our extrapolation method involves generating a sample from the model $\tilde{Y}$ and empirically estimating the frequencies of exceeding high thresholds. The procedure is highly sensitive to the choice of $p_i \in (0,1)$, which determines the threshold $\hat{F}_i^{-1}(p_i)$ in the underlying POT approach. Our goal is to agnostically evaluate the extrapolation power of the model $\tilde{Y}$ depending on $p_i$. We use a martingale testing approach to assess this power and select $p_i$ agnostically. For simplicity, we drop the index $i$ as the same approach applies to each target $i = 1, 2, 3$.

\subsection{Coherence of Top Order Statistics as a Game}
In our pipeline, we use a sample from the model $\tilde{Y}$ to extrapolate from 4 to 50 runs. Since the targets are frequencies of exceeding high thresholds, they depend only on the distributions of the top order statistics $\tilde{Y}_{(k)}$. We omit the dependence on $p$ in the following.

We introduce a game to agnostically test the coherence between the top order statistics from our model $\tilde{Y}$ and the true underlying random variable $Y$. Consider the $K$-largest order statistics $Y_{(K-k)}$ and $\tilde{Y}_{(K-k)}$, for $0 \leq k \leq K-1$, from samples of size $\lceil (1-p) \times 4 \times 165 \times 365 \rceil$. Note that $K$ is fixed and sufficiently small such that $K < (1-p) \times 4 \times 165 \times 365$ for all considered values of $p$.

The game tests the coherence between the extrapolation from top order statistics to the next one in both models. At each round $k$, given the information $\mathcal{F}_k$ (which contains $\sigma(Y_{(K-i)}, \tilde{Y}_{(K-i)}, 0 \leq i \leq k-1)$), we test the null hypothesis:
\[
\mathcal{H}_0: \qquad \mathbb{E}[Y_{(K-k)} \mid \mathcal{F}_k] = \mathbb{E}[\tilde{Y}_{(K-k)} \mid \mathcal{F}_k], \qquad 0\le k\le K-1. 
\]
By construction, the game stops after $K$ steps. The extrapolation power of the model $\tilde{Y}$ for unfeasible steps ($k \geq K$) is assessed by the coherence of the top order statistics over the first $K$ rounds.

\subsection{Testing by Betting Coherence of Top Order Statistics}
We employ a betting strategy to test whether the model $\tilde{Y}$ satisfies the null hypothesis, as in sequential mean hypothesis testing studied in \cite{durand:wintenberger}. The originality lies in considering a sequential game on increasing order statistics rather than on sequentially observed data.

Testing by betting is a powerful tool in sequential games; see \cite{shafer2021testing}. We focus on non-parametric tests for agnosticism and define the Capital test supermartingale as:
\[
L_k^{\rm C}(\gamma_1) = \prod_{i=0}^k \left(1 + (\gamma_1 - 0.5)(\tilde{Y}_{(K-i)} - Y_{(K-i)})\right), \quad k \geq 0,
\]
which is a martingale under $\mathcal{H}_0$ for any $\gamma_1 \in [0,1]$. Note that this procedure for comparing extrapolation models is sensitive to randomness variation in the sample $\tilde{Y}_{(K-i)}$. Alternative strategies, such as those developed in
\cite{Henzi21-comparing}, would address this issue when comparing forecasters. See Section 4.2 of \cite{durand:wintenberger} for a discussion on the advantages of coin betting in this context.

We consider a betting strategy $\gamma_{1,k}$ that is an $\mathcal{F}_k$-adapted process, aiming to predict the sign of the error between the top order statistics from the model and the observed values. The associated wealth process is:
\[
W_k^{\rm C} = \prod_{i=0}^k \left(1 + (\gamma_{1,k} - 0.5)(\tilde{Y}_{(K-i)} - Y_{(K-i)})\right), \quad k \geq 0,
\]
which is also a martingale under $\mathcal{H}_0$. For an efficient betting strategy, the wealth is likely to increase under a well-defined alternative hypothesis. Under the null, no efficient betting strategy exists due to the martingale property.

We choose the Exponential Weighting Averaging (EWA, \cite{cesa}) strategy for betting:
\[
\gamma_{1,k} = \frac{L_{k-1}^{\rm C}(1)}{L_{k-1}^{\rm C}(1) + L_{k-1}^{\rm C}(0)}, \quad k \geq 1, \quad \gamma_{1,0} = 0.5,
\]
for its optimal regret properties against constant strategies $\gamma_1 = 1$ or $0$. Specifically, we have:
\[
\log(W_k^{\rm C}) \geq \max_{\gamma_1 \in \{0,1\}} \log L_k^{\rm C}(\gamma_1) - \log(4), \quad k \geq 0.
\]
Thus, the wealth process is likely to become large if the top order statistics from the model consistently over- or underestimate the largest observations.

\subsection{Power Guarantees for Testing by Betting}
Based on the non-negative martingale property of the wealth process under the null hypothesis $\mathcal{H}_0$, \cite[Theorem 1, p.84]{ville1939} defined the stopping time:
\[
\tau_\alpha^C := \inf_{k \geq 0} \{W_k^C \geq 1/\alpha\},
\]
which is a rejection time for the sequential test satisfying:
\[
\mathbb{P}(\tau_\alpha^C < \infty) \le  \alpha, \quad \text{under } \mathcal{H}_0.
\]

The power guarantees of the testing-by-betting procedure are studied under an appropriate alternative hypothesis. In this non-standard game, the simplest alternative hypothesis introduces a minimal gap $m > 0$:
\[
\mathcal{H}_1: \begin{cases}
\mathbb{E}[\tilde{Y}_{(K-k)} \mid \mathcal{F}_k] \geq \mathbb{E}[Y_{(K-k)} \mid \mathcal{F}_k] + m \text{ for all } k \geq 0, \text{ or} \\
\mathbb{E}[\tilde{Y}_{(K-k)} \mid \mathcal{F}_k] \leq \mathbb{E}[Y_{(K-k)} \mid \mathcal{F}_k] - m \text{ for all } k \geq 0.
\end{cases}
\]

Under this alternative, it takes time for the betting strategy to learn the systematic sign of the conditional error. In games with complex underlying dependencies such as extrapolation, \cite{durand:wintenberger} assumed additionally under the alternative $\mathcal{H}_1$:
\[
|\hat{Y}_{(K-k)} - Y_{(K-k)}| \leq 1, \quad \mathbb{E}[(\hat{Y}_{(K-k)} - Y_{(K-k)})^2 \mid \mathcal{F}_k] \leq m/4, \quad \text{a.s.}, \quad  0\le k\le K-1\,.
\]
They established power guarantees of the form:
\[
\mathbb{E}[\tau_\alpha^C] = \mathcal{O}\left(\frac{\log(1/\alpha)}{m}\right), \quad \text{under } \mathcal{H}_1,
\]
where $\mathcal{O}$ may include logarithmic terms in $\log(1/\alpha)/m$. This rate is optimal for $0 < m < 1$ and $0 < \alpha < 1$. If the errors are deterministic:
\[
\begin{cases}
\mathbb{E}[\tilde{Y}_{(K-k)} \mid \mathcal{F}_k] = \mathbb{E}[Y_{(K-k)} \mid \mathcal{F}_k] + m \text{ for all } k \geq 0, \text{ or} \\
\mathbb{E}[\tilde{Y}_{(K-k)} \mid \mathcal{F}_k] = \mathbb{E}[Y_{(K-k)} \mid \mathcal{F}_k] - m \text{ for all } k \geq 0,
\end{cases}
\]
we are under the alternative, and $\max_{\gamma_1 \in \{0,1\}} \log L_k^{\rm C}(\gamma_1) = (k+1) \log(1 + m/2)$, since the optimal betting strategies are the constant bets $\gamma_1 = 1$ or $0$. Therefore, it takes at least $k$ rounds with:
\begin{equation}\label{eq:confidence}
k+1 \geq \frac{\log(1/\alpha)}{\log(1 + m/2)},
\end{equation}
to reject the null hypothesis. These results are valid in our non-standard game, so martingale testing can be used to agnostically test the coherence of top order statistics from a generative model.

\section{Extrapolation Assessment in Practice}\label{sec:practice}

\subsection{Quantile's Selection Pipeline}
We assess the extrapolation power by evaluating the wealth of the EWA betting strategy at the end of the game after $K=3$ and $5$ steps on a unique simulation $\tilde{y}_{(k)}$ of the top order statistics $\tilde{Y}_{(k)}$ and the largest observations $y_{(k)}$. We want to choose the level $p$ from the grid $\{0.9, 0.99, 0.995, 0.999, 0.9992, 0.9995, 0.9997, 0.9999\}$ that minimizes $W_{K-1}^C$. Note that we did not consider $p=0.9999$ as a possible choice for the level because the associated extrapolation model is estimated on 24 exceedances only. We included it in the grid only for completeness.

We are aware of the possibility of overfitting as we validate the model on the training set due to the lack of top order statistics at the target's level. However the simplicity of the extrapolation model reduces such risk of overfitting on the extremes.

\begin{figure}
\centering
\includegraphics[width=0.4\textwidth]{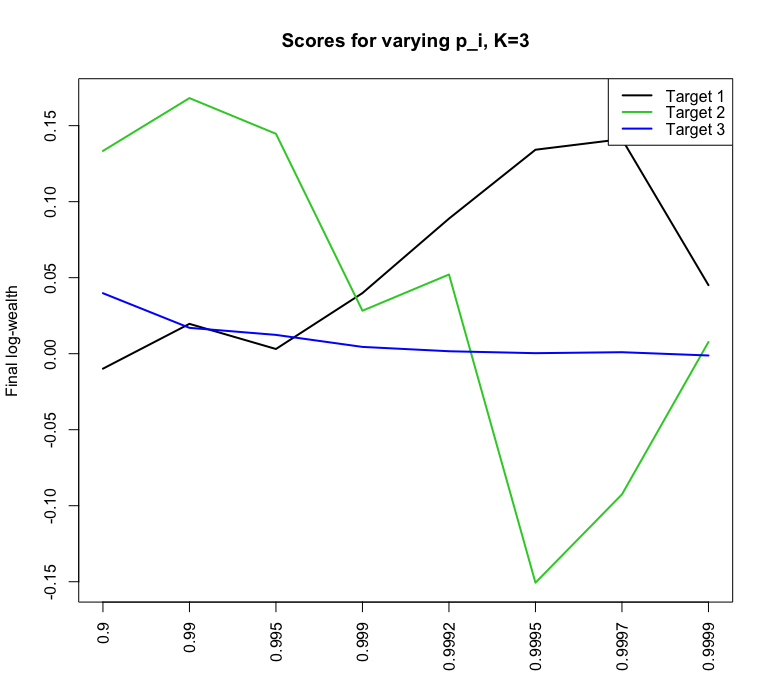}
\includegraphics[width=0.4\textwidth]{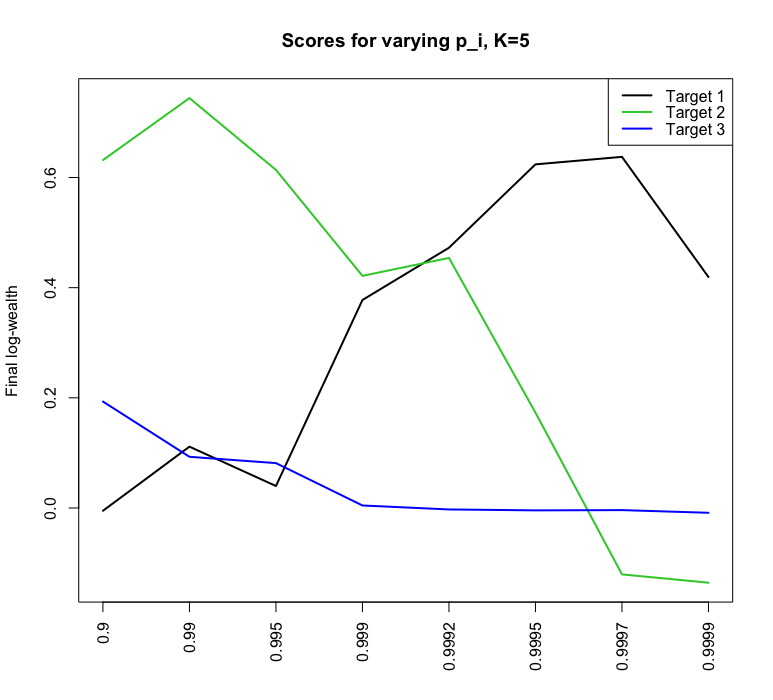}
\caption{Scores obtained for $p_i$ in the the grid $\{0.9, 0.99, 0.995, 0.999, 0.9992, 0.9995, 0.9997, 0.9999\}$ and $K\in\{3, 5\}$.}\label{fig:scores}
\end{figure}

The scores displayed in Figure \ref{fig:scores} are very sensitive to the choice of $K$, the number of rounds in the extrapolation game. Therefore the selection of the level $p$ of the empirical quantile $q$ is not an easy task when it depends on the choice of $K$. Moreover $K$ must be chosen sufficiently small so that large $p$ can be considered.

\subsection{Discussion and Final Choices of Quantiles}
We decided to apply our evaluation to very small values of $K$ because we are interested only in the very top order statistics. Therefore we arbitrarily chose $K=3$ to define the scores.

\begin{itemize}
\item
For Target 2, it yields a clear optimal level $p_2^\star=0.9995$. We therefore selected the associated quantile and frequencies. However, we acknowledge that such a choice of $K$ as 3 does not provide robust theoretical guaranties; using \eqref{fig:scores}, the best betting strategy can only reject same sign errors of predictions of top order statistics larger than $m=0.5$ with confidence $\alpha= 0.5$.
\item For Target 3 the scores are rather insensitive to the level $p$. Excluding $p=0.9999$ as explained above we chose the same level $p_3^\star=0.9995$ as for Target 2.
\item
For Target 1, we observe that the optimal level $p_i^\star$ is very dependent on $K$. For the small $K=3$ our method would have selected $p_i^\star\in\{0.9,0.995\}$. However we were disturbed by the Q-Q plots of our model vs the observation in Figure~\ref{fig:qqplot_target1}. Thus, we chose $p_i=0.999$ graphically based on these Q-Q plots, and did not trust our validation paradigm. In that decision we were also encouraged by the selection of $0.9995$ for Targets 2 and 3. It was a bad choice: as we now know the true frequency was $0.24$, we note that $p_i=0.999$ let us to estimate the frequency to $0.12$, whereas selecting $p_i=0.995$ (resp. $0.9$) would have led us submit $0.26$ (resp. $0.22$) as a point estimate and we would have ranked first on that task.
\begin{figure}
\centering
\includegraphics[width=0.325\textwidth]{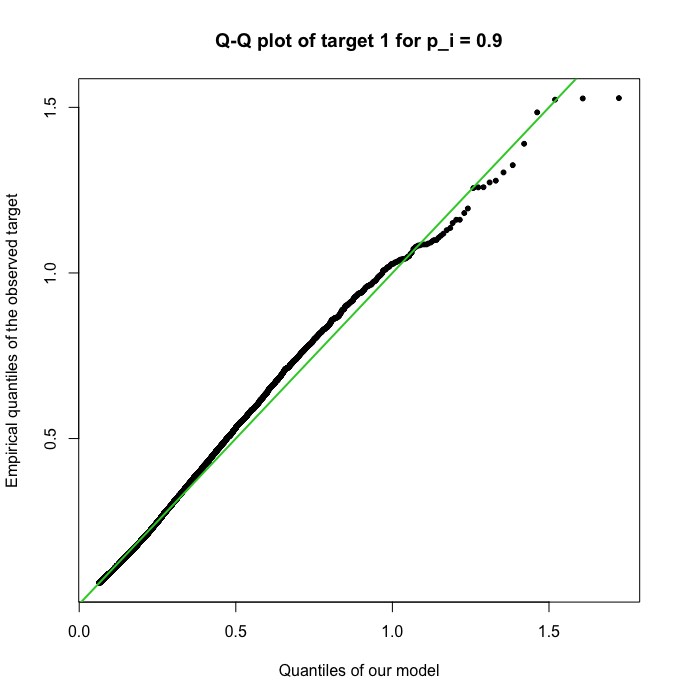}
\includegraphics[width=0.325\textwidth]{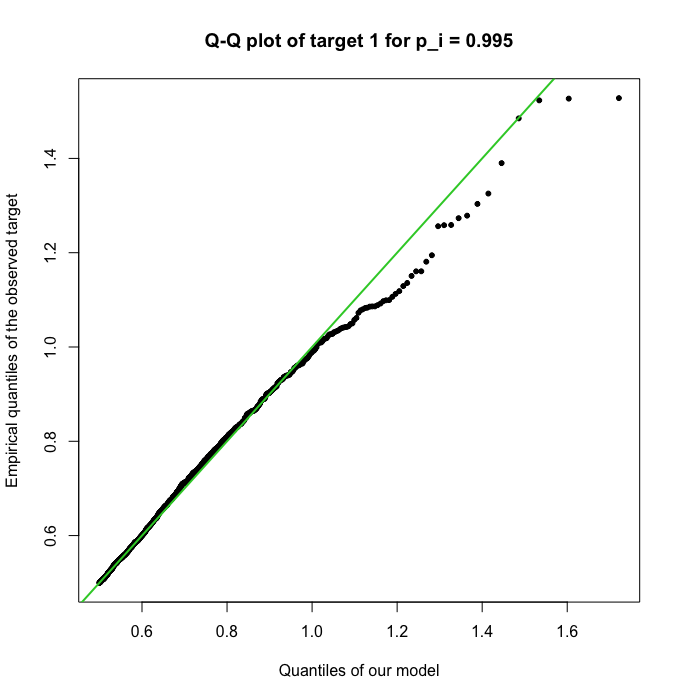}
\includegraphics[width=0.325\textwidth]{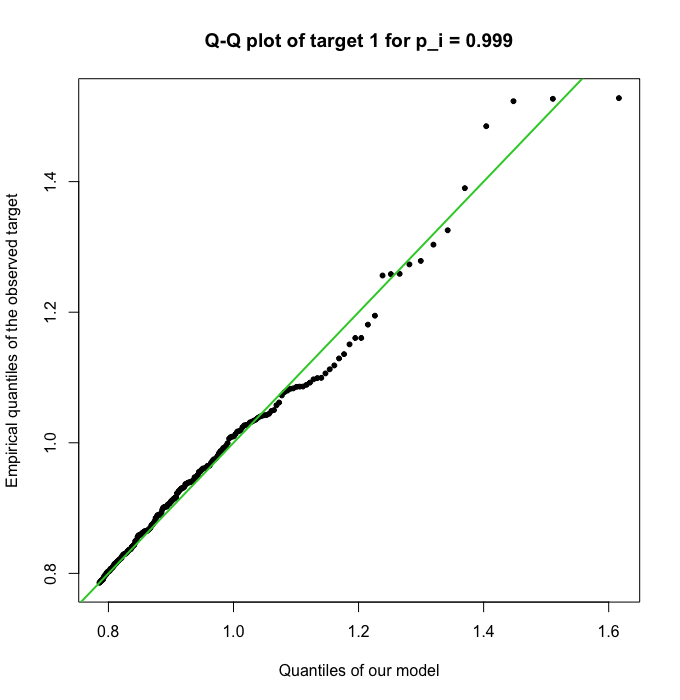}
\caption{Q-Q plots of our model's quantiles vs the observed quantiles for Target 1 and different choices of $p_i$.}\label{fig:qqplot_target1}
\end{figure}

At that time of the competition we were not yet fully aware of the potential danger to use Q-Q plots as diagnostic tools for extrapolation. Indeed the figure represents theoretical quantiles of the model against top-order statistics of the observation. However, even in the simple standard exponential model the difference of the maximum of i.i.d. random variables and the theoretical quantile converges to a Gumbel distribution that is not centered. The positive mean of the Gumbel distribution introduces a bias, and the maximum over-estimates the high quantile of the true distribution.
\end{itemize}



\section{Conclusion and Discussion}
The competition scores were agnostic and coarse. Our POT approach, providing frequencies in multiples of $0.02$ and the Poisson approximation, reducing the range of our confidence intervals, contributed to our winning performance.

A rough univariate analysis like POT proved effective because the scores focused on the excesses of univariate auxiliary targets. The GAM methods used to model the impact of the day of the year were sensitive to large exceedances, particularly when fitted using squared loss in Targets 2 and 3 where our approach was more successful. For these targets, the choice of a "moderately" high probability level $p$ was crucial because it directly impacts the size of the exceedances. We addressed this by the development of agnostic martingale tests on a few top order statistics.

While this approach is novel, it has major limitations such as being intrinsically univariate and lacking rigorous guarantees. Testing by betting relies on rejection times, but the literature lacks clear interpretations of the terminal wealth value, particularly for tests that should be rejected at unfeasible steps $k$ larger than the deterministic stopping time $K-1$. Recent developments on E-values as in \cite{ramdas2024hypothesis} might be useful here. Also, the dependence on $K$ is important and we do not provide a systematic way to choose this parameter in this experimental work. In our extrapolation context, it seems natural to select a very small value to concentrate on top-order statistics close to extrapolation.

Despite these limitations, martingale testing provides unexpectedly high thresholds, which were validated by our scores in the Data Challenge. We won this challenge because we relied on a score that agnostically selected very high quantiles—quantiles we would not have considered, given that very few exceedances (only hundreds) remained to fit a simple seasonal model. In particular, we used GAMs for seasonal adjustment, based on few splines and squared losses, thus highly sensitive to extreme values, yet too simple to overfit. The estimated frequencies are heavily dependent on the selected thresholds. Martingale testing proved sufficiently agnostic to choose the best threshold for extrapolating from such simple models. Further investigations are necessary to evaluate this validation methodology of extrapolation models.

\backmatter

\bmhead{Supplementary information}

Our implementation is available on gitlab: \url{https://gitlab.com/JdeVilmarest/eva2025-data-challenge-viking-sorbonne}.

\bmhead{Acknowledgements}

The authors thank Gloria Buritica for the useful discussions and presenting their approach in full clarity at EVA2025.

\bmhead{Funding}

This research did not receive specific funding.

\begin{appendices}




\end{appendices}


\bibliography{mybib.bib}

\end{document}